\documentclass[12pt, letterpaper]{article}
\pdfoutput=1
\usepackage[left=25.4mm,right=25.4mm,top=25.4mm,bottom=25.4mm]{geometry}
\usepackage[utf8]{inputenc}
\usepackage[version=3]{mhchem} 
\usepackage{siunitx}
\usepackage{graphicx}
\usepackage{amsmath}
\usepackage{authblk}
\usepackage{subcaption}
\begin{document}

\title{Modeling ambient temperature and relative humidity sensitivity of respiratory droplets and their role in Covid-19 outbreaks}
\author[1]{Swetaprovo Chaudhuri \thanks{Email: schaudhuri@utias.utoronto.ca}}
\author[2]{Saptarshi Basu \thanks{Email: sbasu@iisc.ac.in}}
\author[2]{Prasenjit Kabi}
\author[3]{Vishnu R. Unni}
\author[3]{Abhishek Saha \thanks{Email: asaha@eng.ucsd.edu}}
\affil[1]{Institute for Aerospace Studies, University of Toronto, Toronto, Canada}
\affil[2]{Department of Mechanical Engineering, Indian Institute of Science, India}
\affil[3]{Department of Mechanical and Aerospace Engineering, University of California, USA}

\date{}

\maketitle

\begin{abstract}
One of the many unresolved questions that revolves around the Covid-19 pandemic is whether local outbreaks can depend on ambient conditions like temperature and relative humidity. In this paper, we develop a model that tries to explain and describe the temperature and relative humidity sensitivity of respiratory droplets and their possible connection in determining viral outbreaks. The model has two parts. First, we model the growth rate of the infected population based on a reaction mechanism - the final equations of which are similar to the well-known SIR model. The advantage of modeling the pandemic using the reaction mechanism is that the rate constants have sound physical interpretation. The infection rate constant is derived using collision rate theory and shown to be a function of the respiratory droplet lifetime. In the second part, we have emulated the respiratory droplets responsible for disease transmission as salt solution droplets and computed their evaporation time accounting for droplet cooling, heat and mass transfer and finally crystallization of the salt. The model output favourably compares with the experimentally obtained evaporation characteristics of levitated droplets of pure water and salt solution, respectively, ensuring fidelity of the model. Droplet evaporation/desiccation time is indeed dependent on ambient temperature and relative humidity, considered at both outdoor and indoor conditions. Since the droplet evaporation time determines the infection rate constant, ambient temperature and relative humidity are shown to impact the outbreak growth rates.
\end{abstract}




\section*{Introduction}
Seasonality of viral diseases is documented \cite{Deyle13081, jaakkola2014decline, Lowen7692,Cohen2020}, yet its explanation lacks rigorous scientific underpinning. The ongoing Covid-19 pandemic has led to the debate whether temperature and relative humidity have influenced the growth rate of the pandemic in specific geographical locations. The debate has been bolstered by the observation that higher growth rates in localized outbreaks are often associated with higher latitudes and times corresponding to the transition from winter to spring. Wang et al. \cite{wang2020high}
concluded that high temperature and high relative humidity reduced the effective reproductive number of the outbreak - a result they claimed to be consistent with the past outbreaks of influenza and SARS. However, Yao et al. \cite{Yao2000517} found that ambient conditions like temperature do not hold any significant influence on the growth rate of the pandemic. A recent report by the United States National Academies \cite{NAP25771} and references of recent experiments therein show that laboratory evidence from multiple sources suggest reduced survivability of the Covid-19 virus at high temperature and relative humidity. A review of recent literature focusing on establishing effect of temperature and humidity has been conducted in \cite{Mecenas2020.04.14.20064923}. Screening 517 articles the authors found that the general consensus is cold and dry conditions were suitable for the spread of the virus. To our knowledge the reason for this behavior is not well established. 

It has been however well established that Covid-19 transmits via respiratory droplets that are exhaled during breathing, talking, coughing or sneezing. Different activities correspond to different droplet sizes and myriad trajectories for the droplets embedded in the corresponding jets. These respiratory droplets are essentially salt solution droplets with salt mass fraction of about 0.01, in addition to proteins and pathogens, when the droplets emerge from an infected individual \cite{chartier2009natural, xie2007far}. In this paper, to model the outbreaks, we extensively use the evaporation and settling times of NaCl-water droplets which are in turn used as a surrogate model of the infectious droplets. The evaporation mechanism of such droplets is laced with complexities stemming from initial droplet cooling, heat transfer, mass transfer of the solvent and solute, respectively, and finally crystallization of the solute - a phenomenon known as efflorescence. The model developed is first validated with new experimental results obtained from droplets observed to evaporate in an acoustic levitator. Simultaneously, a chemical kinetics based reaction mechanism model is developed with final rate equations similar to that yielded by the SIR model \cite{kermack1927}. The rate constant is shown to be a strong function of the droplet lifetime. Next, the droplet lifetime at two typical locations relevant to the ongoing pandemic is evaluated. At these locations, both outdoor and indoor conditions are considered. Combining all, we find that the model yields encouraging results in at least partially explaining the significantly different growth rate of the outbreak at the two geographical locations exhibiting very different weather conditions. The results in no way suggests that variables not considered in this paper play a secondary role in determining the outbreak spread. Rather, this paper aims to establish a possible connection between the pandemic and the ambient conditions using a well defined framework rooted in physical sciences. The paper is arranged as follows: first we provide details of the experiments used to obtain the evaporation characteristics of the water and salt solution droplets. This is followed by the reaction mechanism model that yields the equations for the growth rate and the infection rate constant of the outbreaks. This infection rate constant provides the connection and motivation for modeling the droplet evaporation time scales. Next, to evaluate the rate constant, detailed modeling of the droplet evaporation is presented. This is followed by results and discussions. Finally, we summarize the approach and findings in the conclusion section.

\section*{Experiments}
\begin{figure}[ht]
\begin{centering}
\includegraphics[width=0.6\textwidth]{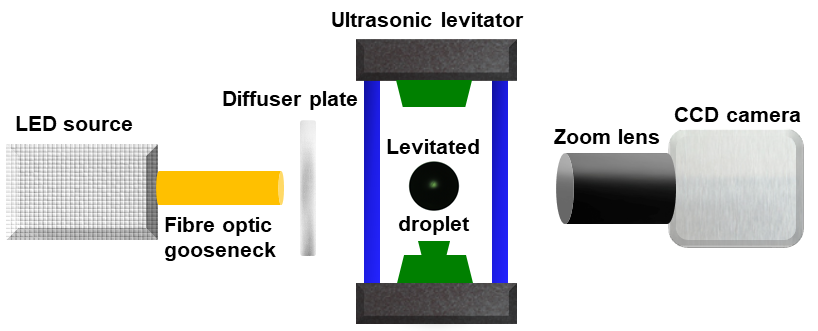}

\caption{Experimental setup showing the acoustic levitation of a droplet illuminated by a cold LED source. A diffuser plate is used for uniform imaging of the droplet. A CCD camera fitted with zoom lens assembly is used for illumination. The schematic is not to scale.}
\label{Fig:Levitator}
\end{centering}
\end{figure}

The experiments with isolated evaporating droplets were conducted in a contact-less environment of an ultrasonic levitator (tec5) to discount boundary effects, generally present in suspended, pendant or sessile droplet setups \cite{saha_JFM, saha_IJHMT}. The experimental setup with the diagnostics is shown in Fig. \ref{Fig:Levitator}.
A droplet was generated and positioned near one of the stable nodes of the levitator by using a micropipette. The levitated droplet was allowed to evaporate in the ambient condition of the lab  at 30$^o$C and at about 50\% RH. The transient dynamics of evaporation and precipitation of the evaporating droplet was captured with shadowgraphy technique using combination of a CCD camera (NR3S1, IDT Vision) fitted with a Navitar zoom lens assembly (6.5x lens and 5x extension tube) and a backlit-illumination by a cold LED light source (SL150, Karl Storz). 

A set of 10 images at a burst speed of 30 frames per second is acquired every 2 seconds for the entire duration of the droplet lifetime. The spatial resolution of the images was $\approx 1\mu$m/pixel. The temporal evolution of diameter of the evaporating droplet was extracted from the images using the "Analyze Particles" plugin in ImageJ (open source platform for image processing). The final precipitate was carefully collected on carbon tape and observed in dark-field mode under a reflecting microscope (Olympus BX-51). A range of initial droplet diameters varying from $300\mu m$ to $1000\mu m$ were investigated in experiments.      

\section*{A reaction mechanism to model the pandemic}
\begin{figure}[ht]
\includegraphics[width=0.65\textwidth]{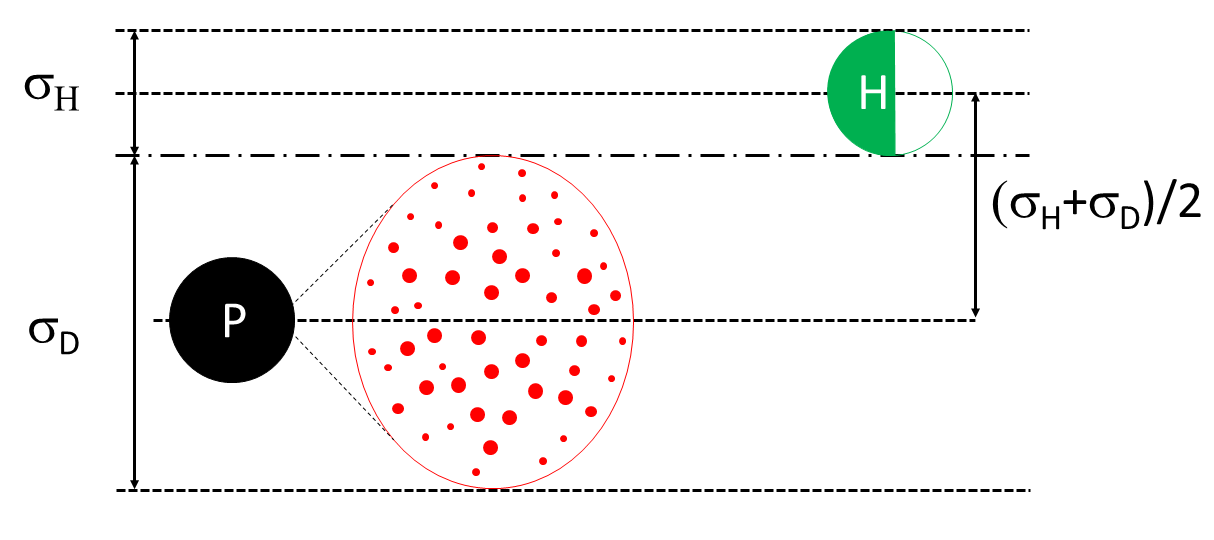}
\begin{centering}
\caption{A schematic of the collision rate model for the infection to occur. Infected person $P$ (large black circle) ejects a cloud of infectious droplets $D$ denoted by small red dots and the cloud approaches a healthy person $H$ with a relative velocity $\vec{V}_{DH}$ to infect them. The figure also shows the collision volume swept by the droplet cloud $D$ and $H$ with their respective effective diameters.}
\label{Fig:CK}
\end{centering}
\end{figure}

In this section, we model the infection spread rate using the collision theory of reaction rates, well known in chemical kinetics \cite{law_2006}. The connection between droplets and the outbreak will be established later. In this model we adopt the following nomenclature: $P$ represents a Covid-19 positive person infecting healthy person(s) susceptible to infection. The healthy person is denoted by $H$ (who is initially Covid-19 negative), $R$ represents a person who has recovered from Covid-19 infection and hence assumed to be immune from further infection while $X$ represents a person who dies due to Covid-19 infection. We consider only one-dimensional head on collisions and the schematic of a collision volume is shown in Fig. \ref{Fig:CK}. Here, one healthy person denoted by $H$ with effective diameter $\sigma_H$ is approached by a Covid-19 positive person $P$ of same effective diameter. $\sigma_H$ can be considered as the diameter of the hemispherical volume of air that is drawn by $H$ during each act of inhalation. 

It is widely believed that Covid-19 spreads by respiratory droplets \cite{CDC} resulting from breathing, coughing, sneezing or talking. Thus, we assume that a conical volume in front of $P$ is surrounded by a cloud of infectious droplets exhaled by $P$. The droplet cloud is denoted by $D$ and the maximum cloud diameter is given by $\sigma_D$. Clearly $\sigma_D$ should be determined by the smaller of the evaporation or settling time of the droplets ejected by $P$ and the horizontal component of the velocity with which the droplets displace. In each such cloud, we assume that there are numerous droplets containing active Covid-19 virus. The velocity of this droplet cloud relative to $H$ is given by $\vec{V}_{DH}$. In such a scenario, we assume that in a unit volume there are $n_P$ infected persons, $n_H$ healthy persons. For a collision to be possible, the maximum separation distance between the centers of $D$ (the droplet cloud) and $H$ is given by 
\begin{equation} \label{Eqn.collision diameter}
\sigma_{DH}=(\sigma_D + \sigma_H)/2
\end{equation}

The collision volume - the volume of the cylinder within which a collision between the droplet cloud of $P$ and air collection volume of $H$ should lie is given by $\pi \sigma_{DH} ^2 V_{DH}$.
Thus, the number of collisions between $H$ and the droplet cloud $D$ of $P$, per unit time per unit volume, that will trigger infections, is given by 
\begin{equation} \label{ZDH}
Z_{DH}=\pi \sigma_{DH} ^2 V_{DH} n_P n_H
\end{equation}
Now, given that each collision between $P$ (basically its droplet cloud $D$) and $H$ results conversion of the healthy individual to infected individual, we can write
\begin{equation} \label{dnH}
\frac{dn_H}{dt}= -Z_{DH}
\end{equation}

Now, we can define $[P]=n_P/n_{total}$ and $[H]=n_H/n_{total}$, whereas $n_{total}$ is the total number of people per unit volume in motion outside their residence - those who are capable of transmitting the infection as well as accepting the infection, in that given volume. This implies

\begin{equation} \label{Eqn:omega}
\omega = -\frac{d[H]}{dt}=  n_{total} \pi \sigma_{DH} ^2 V_{DH} [P][H] = k[P][H] 
\end{equation}

where, 
\begin{equation} \label{Eqn:k}
k =  n_{total} \pi \sigma_{DH} ^2 V_{DH}
\end{equation}

Here, $\omega$ is the reaction rate.
Furthermore, if we assume that the mortality rate is about 3$\%$ for the ongoing Covid-19 pandemic, then we can convert the kinetics of infection spread to a complete reaction mechanism given by the following:

$$\ce{ \textit{P} + \textit{H} ->[$k_1$] \textit{P} + \textit{P}^* } \ \ \ \ \ \ \ \ [R1] $$
$$\ce{ \textit{P}^* ->[$k_2$] \textit{P}  } \ \ \ \ \ \ \ \ \ \ \ \ \ \ \ \ \ \ \ \ [R2] $$
$$\ce{ \textit{P} ->[$k_3$] 0.97\textit{R} + 0.03\textit{X}  } \ \ \ \ [R3] $$

It is to be recognized that $H$ does not become $P$ immediately on contact with the droplet cloud. 
The virus must proliferate for a finite time after contact to render a person infectious. A person who has just come in contact with the virus and does not have the capability to infect others yet, is denoted by $P^*$.
$k_1$, $k_2$ and $k_3$ are the rate constants of reactions [R1], [R2] and [R3], respectively. 
All rate constants must have dimensions of $[T]^{-1}$. Clearly, $k_1 > k_3$ for the rapid outbreak to occur. It is to be recognized that this framework implies that $k_1$, the rate constant of the second order elementary reaction $[R1]$ resulting from collisions between the droplet cloud from an infectious individual and healthy individual, is purely controlled by physical effects. The rate constants $k_2$ and $k_3$ of the other two first order elementary reactions $[R2]$ and $[R3]$ are essentially decay rates emerging from the time by which the respective concentrations reach $e^{-1}$ levels of the initial concentration, for the respective reactions. Thus, $k_2$ and $k_3$ are purely determined by interaction between the virus and human body. We know the approximate recovery time from the Covid-19 disease is about 14 days. Thus we can assume $k_3 = 1/14$ day$^{-1}$. We also assume the latency period (not incubation period) to be 1 day, hence $k_2=1$ day$^{-1}$. Given the importance of $k_1$ in determining the outbreak characteristics, we will refer to $k_1$ as the infection rate constant. The major contribution of the work is imparting a rigorous physical interpretation to $k_1$ and calculating it ab initio.

Using Eqn. \ref{Eqn:omega}, we can write the system of ODEs for $d[P]/dt$ and $d[P^*]/dt$ as

\begin{equation} \label{Eqn:dPdt0}
\begin{bmatrix}
\dfrac{d[P]}{dt} \\ \\ \dfrac{d[P^*]}{dt}
\end{bmatrix} = \begin{bmatrix} -k_3 & k_2 \\ \\ k_1[H] & -k_2 \end{bmatrix} 
\begin{bmatrix} 
[P] \\ \\ [P^*] 
\end{bmatrix}
\end{equation}

In this paper we are interested in modeling the initial phases of the outbreaks where $[H] \gg [P]$. Hence, we can safely assume $[H]\approx[H]_0 $ i.e. the concentration of healthy people remains approximately constant during the early phase of the outbreak and is equal to the initial concentration which is very close to unity at $t=0$ i.e. at the onset of the outbreak. The time of the beginning of the outbreak denoted by $t=0$ for a particular location can be assumed to be the day when number of Covid-19 positive persons equalled 10. $[P]_0$ is $[P]$ at $t=0$.
Then, $[P]$ can be solved as an eigenvalue problem and is given by 

\begin{equation} \label{P=f(t)}
[P]=[P]_0(C_1e^{\lambda_1t} + C_2e^{\lambda_2t})
\end{equation}

$C_1, C_2$ are constants to be determined from the eigenvectors and the initial conditions $[P]_0$ and $[P^*]_0$. $\lambda_{1,2}$ are the eigenvalues and are given by
\begin{equation} \label{lambda}
\lambda_{1,2} = \frac{-(k_3 + k_2) \pm \sqrt{(k_3 + k_2)^2 - 4(k_2k_3 -k_1k_2)}}{2}
\end{equation}
 By Eqn. \ref{Eqn:k}, $k_1 = n_{total} \pi \sigma_{DH} ^2 V_{DH}$. As mentioned before $k_2=1$ day$^{-1}$ and $k_3=1/14$ day$^{-1}$ which yields $\lambda_{1,2} = -0.5357 \pm \sqrt{0.2156 + k_1}$. If $k_2\rightarrow\infty$ i.e. a healthy person becomes infectious immediately on contact with an infectious person $\lambda_1 \rightarrow k_1-k_3$.
 
Clearly this model does not yet account for the preventive measures like "social distancing", "quarantining" after contact tracking and population wide usage of masks. We will call this "social enforcement". However, it can be included by accounting for the time variation of $[H]$. Social enforcement reduces the concentration of healthy, susceptible individuals from $[H_0]$ to $[H_{SE}]$ where the concentration of healthy population susceptible to infection after implementing strict social distancing (at time $t=t_{SE}$) $[H_{SE}] < [H_0]$. In case of social enforcement $[P]$ will be given by: 
\begin{equation}
\begin{split}
[P]=[P]_0(C_1e^{\lambda_1t} + C_2e^{\lambda_2t}) \quad 0<t<t_{SE} 
\\
[P]=[P]_{SE}(C_1e^{\lambda_1(t-t_{SE})} + C_2e^{\lambda_2(t-t_{SE})})   \quad t\geq t_{SE}
\end{split}
\label{Eq:Social Enforcement}
\end{equation}
Here, $[P] = [P]_{SE}$ at $t=t_{SE}$ and $\lambda_{1,SE}, \lambda_{2,SE}$ are the eigenvalues from Eqn. \ref{Eqn:dPdt0} with $[H]=[H_{SE}]$. $k_1$ - the infection rate constant, remains to be completely determined. It is to be recognized that two of the key inputs of $k_1$ are $\sigma_{DH}$  and $V_{DH}$, since $k_1 \propto V_{DH}\sigma_{DH}^2$ by Eqn. \ref{Eqn:k}. As already mentioned $\sigma_H$ is the diameter of the hemisphere from which breathable air is inhaled. $\sigma_D$ is the diameter of the droplet cloud. Using mass and momentum balance, it can be shown that the diameter of a turbulent puff initially injected with a velocity $U_0$, after time $t$ is given by $\sigma_{D,t} = \sigma_{D,0}(8aU_0t/(9\sigma_{D,0}))^{1/4}$ \cite{Cushman2008}. The velocity of the puff, on the other hand, can be expressed as, $V_{D,t} = U_0(9\sigma_{D,0}/(8aU_0t))^{3/4}$ \cite{Cushman2008}. Since the puff occurs only when $P$ coughs/sneezes, the effective, time averaged diameter and the velocity of the puff relative to $P$ can be calculated as 

\begin{equation}
\sigma_{D,P} = {\tau}^{-1}\int_{0}^{\tau} \sigma_{D,t} dt =  \frac{4\sigma_{D,0}}{5}\left(\frac{8 a U_0}{9\sigma_{D,0}} \right)^{1/4} \tau^{1/4}
\label{Eq:sigma_D}
\end{equation}

\begin{equation}
V_{D,P} = {\tau}^{-1}\int_{0}^{\tau} V_{D,t} dt =  4U_0\left(\frac{9\sigma_{D,0}}{8aU_0}\right)^{3/4} \tau^{-3/4}
\label{Eq:V_d}
\end{equation}



Here, $\tau$ is the characteristic lifetime of the respiratory droplets in the puff, $\sigma_{D,0}$ is the initial diameter of the jet, entrainment coefficient $a=2.25$ \cite{Cushman2008}. 
Now, the puff is present only for a short time $\tau$ after it has been ejected. Therefore the steady state $k_1$ can be defined as

\begin{equation}
k_1 =  n_{total} \pi \sigma_{DH} ^2 V_{DH} (\tau / t_c)
\label{Eq:k_1 equation}
\end{equation}
Just like in collision theory, not all molecules are energetic enough to effect reactions, in our case the droplet cloud is not always present. The last fraction $(\tau / t_c)$ is the probability the droplet cloud with average diameter $\sigma_{D,P}$ is present. $t_c$ is the average time period between two vigorous expiratory events. $V_{DH}=(V_{D,P} + V_P) + V_H$. We can assume $V_P = V_H$. It is thus apparent that $\tau$ appears in $\sigma_{DH}$, $V_{DH}$ and in the last fraction in Eqn. \ref{Eq:k_1 equation} thereby emerging as a critical parameter of the entire pandemic dynamics. Hence $\tau$ merits a detailed physical understanding.
Given the composition of the respiratory droplets, modeling $\tau$ is highly non-trivial and is taken up in the following section.

\section*{Modeling respiratory droplet evaporation}
It is well documented in the literature that an average human exhales droplets (consisting of water, salt, proteins and virus/bacteria) in the range of $1-100\mu m$ \cite{duguid1945numbers, xie2009exhaled, xie2007far}. In this section, we offer a detailed exposition of the evaporation dynamics of such droplets as ejected during the course of breathing, talking, sneezing or coughing.


The small droplets ($<2-3 \mu m$) have a very short evaporation timescale. This implies that these droplets evaporate quickly ($<1 s$) after being breathed out. However, the same conclusion does not hold for slightly larger droplets ejected in form of cloud or aerosol ($> 5 \mu m$). These droplets exhibit longer evaporation time, leading to increased chances of transmission of the droplet laden viruses. 
In particular, when inhaled, these droplets enable quick and effective transport of the virus directly to the lungs airways causing higher probability of infection. As a rule of thumb, the smaller droplets ($<30 \mu m$) have low Stokes number, thereby allowing them to float in ambient air without the propensity to settle down. For larger droplets ($>100 \mu m$), the settling timescale is very small ($\sim 0.5 s$). In effect, based on the diameter of the exhaled droplets, we propose three distinct possibilities:
\begin{itemize}
    \item Small droplets ($<5 \mu m$) evaporate within a fraction of a second, thereby reducing the possibility of infection.
    \item Large droplets ($>100 \mu m$) settle within a small time frame ($<0.5s$) limiting the radius of infection.
    \item Intermediate droplets ($\sim 30 \mu m$) show the highest probability of infection due to slightly longer evaporation lifetime and low Stokes number. The significance of the $30 \mu m$ diameter droplets will be apparent later.
\end{itemize}

In this work, we particularly focus our attention to the modeling of droplets over a large range of diameters from $1\mu m$ to $100 \mu m$. Based on available literature, we assume that the droplets exhaled during breathing, are at an initial temperature of $30^o C$ \cite{CARPAGNANO2017855}. The ambient condition, however, vary strongly with geographical, seasonal changes etc. Hence, in the following, we conduct a parametric study to determine the droplet lifetime across a large variation of temperature and relative humidity conditions. The droplet evaporation physics is complicated by the presence of non-volatile salts (predominantly NaCl) as present in our saliva \cite{chartier2009natural}. We would also look into simultaneous desiccation of solvent and crystallization of such salts in subsequent subsections. Once the exhaled droplet encounters ambience, the droplet will evaporate as it undergoes simultaneous heat and mass transfer. 
\subsection*{Evaporation} 
For modeling purpose, the exhaled droplets are assumed to evaporate in a quiescent environment at a fixed ambient temperature and relative humidity. In reality, during coughing, talking or sneezing, the droplets are exhaled in a turbulent puff \cite{bourouiba2014violent}. However, as shown in Eqn. \ref{Eq:V_d}, the puff rapidly decelerates due to entrainment and lack of sustained momentum source, rendering the average $V_{D,P}$ to be less than 1\% of the initial velocity. Furthermore, since Prandtl number $Pr=\nu/\alpha =0.71$ for air, we can safely assume that the temperature and relative humidity that the droplets in the puff experiences are on average very close to that of the ambient.
At the initial stages, the puff will indeed be slightly affected by buoyancy, which will influence droplet cooling and evaporation dynamics. Quantifying these effects accurately, merit separate studies, see for e.g. \cite{Narasimha16164} for buoyant clouds. In a higher dimensional model these could be incorporated. In any case, the evaporation rate of the droplet is driven by the transport of water vapor from the droplet surface to the ambient far field. Assuming quasi-steady state condition, the evaporation mass flux can be written as 
\begin{equation}
\begin{split}
    \dot{m}_1=-4\pi\rho_v D_v R_s log(1+B_M) \\
    \dot{m}_1=-4\pi\rho_v \alpha_g R_s log(1+B_T)
\end{split}
    \label{eq:mdot}
    \end{equation}
here, $\dot{m}_1$ is the rate of change of droplet water mass due to evaporation, $R_s$ the instantaneous droplet radius, $\rho_v$ is density of water vapor and $D_v$ is the binary diffusivity of water vapor in air, $\alpha_g$ is the thermal diffusivity of surrounding air. $B_{M}=(Y_{1,s}-Y_{1,\infty})/(1-Y_{1,s})$ and $B_{T}=C_{p,l}(T_{s}-T_{\infty})/h_{fg}$ are the Spalding mass transfer and heat transfer numbers, respectively. Here, $Y_1$ is mass fraction of water vapor, while subscript $s$ and $\infty$ denote location at droplet surface and at far field, respectively. The numerical subscripts 1, 2 and 3 will denote water, air and salt respectively. $C_{p,l}$ and $h_{fg}$ are the specific heat and specific latent heat of vaporization of the droplet liquid. For pure water droplet, the vapor at the droplet surface can be assumed to be at the saturated state. However, as indicated earlier, the exhaled droplets during talking, coughing or sneezing are not necessarily pure water, rather they contain plethora of dissolved substances \cite{xie2007far}. The existence of these dissolved non-volatile substances, henceforth denoted as solute, significantly affects the evaporation of these droplets by suppressing the vapor pressure at the droplet surface. The modified vapor pressure at the droplet surface for binary solution, can be expressed by Raoult's Law, $P_{vap}(T_s,\chi_{1,s})=\chi_{1,s}P_{sat}(T_s)$, where $\chi_{1,s}$ is the mole fraction of evaporating solvent (here water) at droplet surface in the liquid phase \cite{Sirignano_2010} and $\chi_{1,s} = 1-\chi_{3,s}$. The far field vapor concentration, on the other hand, is related to the relative humidity of the ambient.
Considering the effects of Raoult's law and relative humidity, the vapor concentrations at droplet surface and at far field can be expressed as: 
\begin{equation} \label{eq:Yfs}
\begin{split}
Y_{1,s}=\frac{P_{vap}(T_s, \chi_{1,s})M_1}{P_{vap}(T_s,\chi_{1,s})M_1 + (1-P_{vap}(T_s,\chi_{1,s}))M_2}  \\
Y_{1,\infty}=\frac{(RH) P_{sat}(T_{\infty})M_1}{(RH) P_{sat}(T_{\infty})M_1 + (1-(RH) P_{sat}(T_{\infty}))M_2} 
\end{split}
\end{equation}

For evaporation, the droplet requires latent heat, which is provided by droplet's internal energy and surrounding ambient. Assuming perfect stirring within the droplet (no thermal gradients), the energy balance is given by
\begin{equation} 
    mC_{p,l}\frac{\partial T_s}{\partial t}=-k_gA_s\frac{\partial T_s}{\partial r}|_s + \dot{m}_1 h_{fg} - \dot{m}_1 e_l
    \label{eq:HT}
\end{equation}
where, $T_s$ is instantaneous droplet temperature; $m=(4/3)\pi\rho_l R_s^3$ and $A_s=4\pi R_s^2$ are the instantaneous mass and surface area of the droplet; $\rho_{l}$ and $e_l$ are the density and specific internal energy of the binary mixture of salt (if present) and water and $k_g$ is the conductivity of air surrounding the droplet. $\frac{\partial T}{\partial r}|_s$, is the thermal gradient at the droplet surface and can be approximated as $(T_s-T_\infty)/R_s$, which is identical to convective heat transfer for a sphere with Nusselt number of 2.

\subsection*{Crystallization}
Evaporative loss of water leads to increase in the salt concentration in the droplet, with time. As shown before $P_{vap}(T_s,\chi_{1,s})$ is a function of the salt concentration in the droplet which thus must be modeled using the species balance equation as shown below in Eqn. \ref{eq:Salt mass fraction}

\begin{equation} 
    \frac{d mY_3}{d t} + \dot{m}_{3,out} = 0
    \label{eq:Salt mass fraction}
\end{equation}

$\dot{m}_{3,out}$ represents the rate at which salt mass leaves the solution due to crystallization and is modeled below. Clearly the model shows that as water leaves the droplet, $Y_3$ increases. When $Y_3$ is sufficiently large such that the supersaturation ratio $S = Y_3/Y_{3,c}$ exceeds unity, crystallization begins. Here we use $Y_{3,c}=0.393$ based on the efflorescent concentration of 648 g/L reported for NaCl-water droplets in \cite{gregson2018drying}. Growth rate of the crystal could be modelled using a simplified rate equation from \cite{naillon2015evaporation, derluyn2012salt}. 

\begin{equation} 
    \frac{d l}{d t} = (S-1)^{g_{cr}}C_{cr}e^{-E_a/RT_s}
    \label{Eqn:Crystal growth rate}
\end{equation}

Here $l$ is the crystal length. By \cite{derluyn2012salt}, for NaCl, the constant $C_{cr}=1.14 \times 10^4 m/s$, the activation energy $E_a = 58180 J/mol$  and a constant $g_{cr} = 1$.
Using this, the rate of change of the crystal mass which equals $\dot{m}_{3,out}$, is given by \cite{derluyn2012salt}

\begin{equation} 
    \dot{m}_{3,out} = \frac{d m_{3,crystal}}{d t} = 6 \rho_s (2l)^2  \frac{d l}{d t}
    \label{eq:Crystal mass}
\end{equation}


The governing equations (Eqns. \ref{eq:mdot}-\ref{eq:Crystal mass}) manifest that several physical mechanisms are coupled during the evaporation process. The internal energy of the droplet undergoes variation primarily due to evaporative mass loss and sensible enthalpy change, resulting in change in its temperature. The evaporation is driven by the vapor concentration gradient from the droplet surface to the far field. The partial pressure of water vapor at the droplet surface is determined by the surface temperature and solvent (water) mole-fraction at the droplet surface, while its surface temperature $T_s$ is determined by Eqn. {\ref{eq:HT}}. The far field vapor concentration, on the other hand, is ascertained by the ambient condition, (temperature and relative humidity, $RH$), which in the case of Covid-19 corresponds to specific geographical locations. It is obvious that for lower ambient temperatures, i.e. $T_{s,0}>T_{\infty}$, the droplet should undergo rapid cooling from its initial value. The droplet temperature, however, should eventually reach a steady state limit (wet bulb). This limit is such that the droplet surface temperature will be always lower than the ambient, implying a positive temperature gradient or heat input. The heat subsequently transferred from the ambient to the droplet surface after attaining wet bulb limit is used completely for evaporating the drop without any change in sensible enthalpy. For a droplet with pure water, i.e. no dissolved non-volatile content, the mole-fraction of solvent at surface remain constant at 100$\%$ and at the limit of steady state, the droplet evaporation can be written in terms of the well-known $D^2$ law \cite{law_2006,Sirignano_2010}
\begin{equation} \label{Eqn:tau_d_m}
D^2_s(t)=D^2_{s,0}-K_mt
\end{equation}
where,
\begin{equation} \label{Eqn:K_evap}
K_m = 8(\rho_v/\rho_l)D_wln (1 + B_{M}).
\end{equation}

However, for a droplet with binary solution, the evaporation becomes strongly dependent on the solvent (or solute) mole-fraction, which reduces (or increases) with evaporative mass loss. The transient analysis, thus, becomes critically important in determining the evolution of droplet surface temperature and instantaneous droplet size. During evaporation, the mole-fraction of solute increases and attains a critical super-saturation limit, which triggers precipitation. The precipitation and accompanied crystallization dynamics, essentially, reduce the solute mass dissolved in liquid phase leading to a momentary decrease in its mole-fraction. This, in turn increases the evaporation rate as mandated by Raoult’s law, which subsequently increases the solute concentration. These competing mechanisms control evaporation at the latter stages of the droplet lifetime. At a certain point, due to continuous evaporation, the liquid mass completely depletes and evaporation stops. The droplet after complete desiccation consists only of salt crystals, probably encapsulating the viruses and rendering them inactive. 

\section*{Results and discussions}
\subsection*{Experimental validation}
\begin{figure}[ht]
\includegraphics[width=0.85\textwidth]{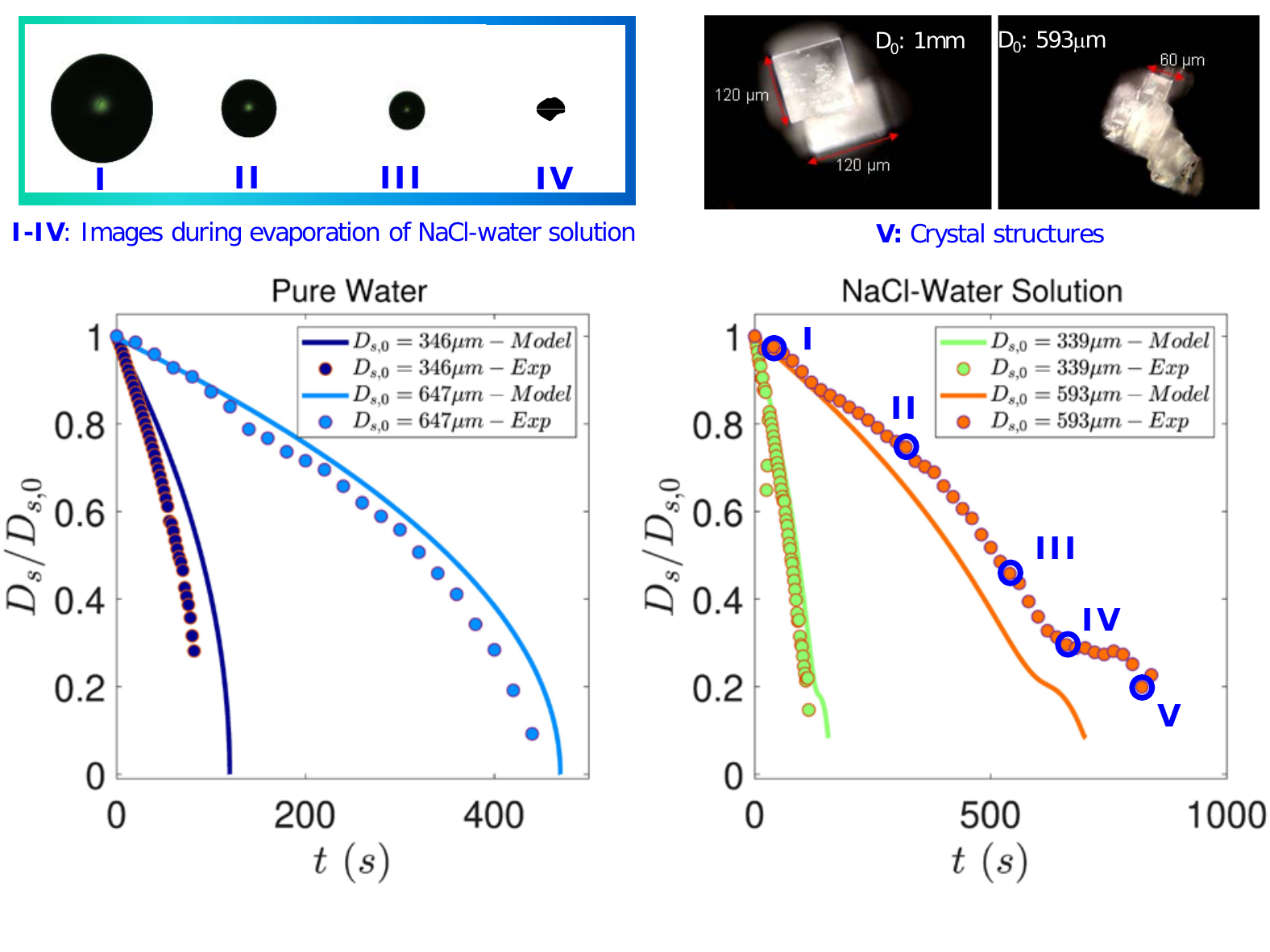}
\begin{centering}
\caption{ Instantaneous droplet images taken by CCD camera (top left panel) and Dark field micrograph of the final salt precipitate (top right panel). Comparison of experiments and simulations in the bottom left and right panels. Evolution of normalized droplet diameter as a function of time for pure water (left panel); salt-water solution droplet with 1\% NaCl (right panel).}
\label{Fig:Exp_comparison}
\end{centering}
\end{figure}

To validate the model, few targeted experiments were conducted to observe  isolated levitated droplets evaporating in a fixed ambient condition.  Particularly, the droplets with (1\% w/w) NaCl solution vaporized to shrink to 30\% of its initial diameter during the first stage of evaporation as shown in Fig. \ref{Fig:Exp_comparison}.  Hereafter, a plateau-like stage is approached due to increased solute accumulation near the droplet’s surface, which inhibits the diameter from shrinking rapidly. However, as shown in Fig. \ref{Fig:Exp_comparison} shrinkage does occur (till $D_s/D_{s,0} \approx 0.2$), as the droplet undergoes a sol-gel transformation. The final shape of the precipitate is better observed from the micrographs presented in Fig. \ref{Fig:Exp_comparison}. 

Figure \ref{Fig:Exp_comparison} shows the final precipitate morphology for the desiccated droplets. The precipitates display a cuboid shaped crystalline formation which is consistent with the structure of NaCl crystal. The size and crystallite structure does show some variation which could be linked with the initial size of the droplet. Only precipitates from larger droplets could be collected since smaller sized precipitates tend to de-stabilize and fly-off the levitator post-desiccation. While the precipitate from larger sized droplets tend to yield larger and less number of crystals, smaller droplets seem to degenerate into even smaller crystallites. However, this work does not investigate the dynamics of morphological changes of crystallization in levitated droplets.  

Figure \ref{Fig:Exp_comparison} shows the comparison between results obtained from experiments and modeling. Experiments were performed with both pure water droplets as well as with droplets of 1\% salt solutions. Experiments have been described in the previous section. For the pure water cases shown in the left panel, simulation results follow the experiments rather closely.
In the pure water case, classical $D^2$ law behavior could be observed. For the salt water droplets, a deviation from the $D^2$ law behavior occurs and droplet evaporation is slowed. This is due to the reduced vapour pressure $P_{vap}(T_s, \chi_{1,s})$ on the droplet surface resulting from the increasing salt concentration with time. Evaporation rate approaches zero at about $D_s/D_{s,0}$ for 0.3 (experiments) and 0.25 (simulations), respectively. However, the salt concentration attained at this stage exceeds the supersaturation $S \geq 1$ required for onset of crystallization. Thus the salt crystallizes reducing its concentration and increasing $P_{vap}(T_s, \chi_{1,s})$ such that evaporation and water mass loss can proceed until nearly all the water has evaporated and only a piece of solid crystal as shown in Fig. \ref{Fig:Exp_comparison} is left.
It can be observed from Fig. \ref{Fig:Exp_comparison} that in all cases, the final evaporation time is predicted within 15\% of the experimental values. This suggests that despite the model being devoid of complexities associated with inhomogeneities of temperature and solute mass fraction within the droplet and simple one step reaction to model the crystallization kinetics, the model demonstrates reasonably good predictive capability. It is prudent to mention again, that although we have done the analysis for single isolated droplet, in reality coughing or sneezing involves a whole gamut of droplet sizes in the form of a cloud. However, based on the droplet size distribution during coughing and sneezing \cite{duguid1946size}, one can show that the total volume of the respiratory droplets is a negligible fraction ($O(10^{-5})$) of the volume that an individual exhales during breathing or coughing \cite{bourouiba2014violent}. This ensures that the droplets are well dispersed and the gas is sufficiently dilute. In other words, the dilute limit dictates that each droplet in that exhaled volume may not be influenced by the other droplets in the vicinity. This further justifies the efficacy of the current model based on an isolated droplet.

\begin{figure}[h!]
\centering
\includegraphics[width=0.6\linewidth]{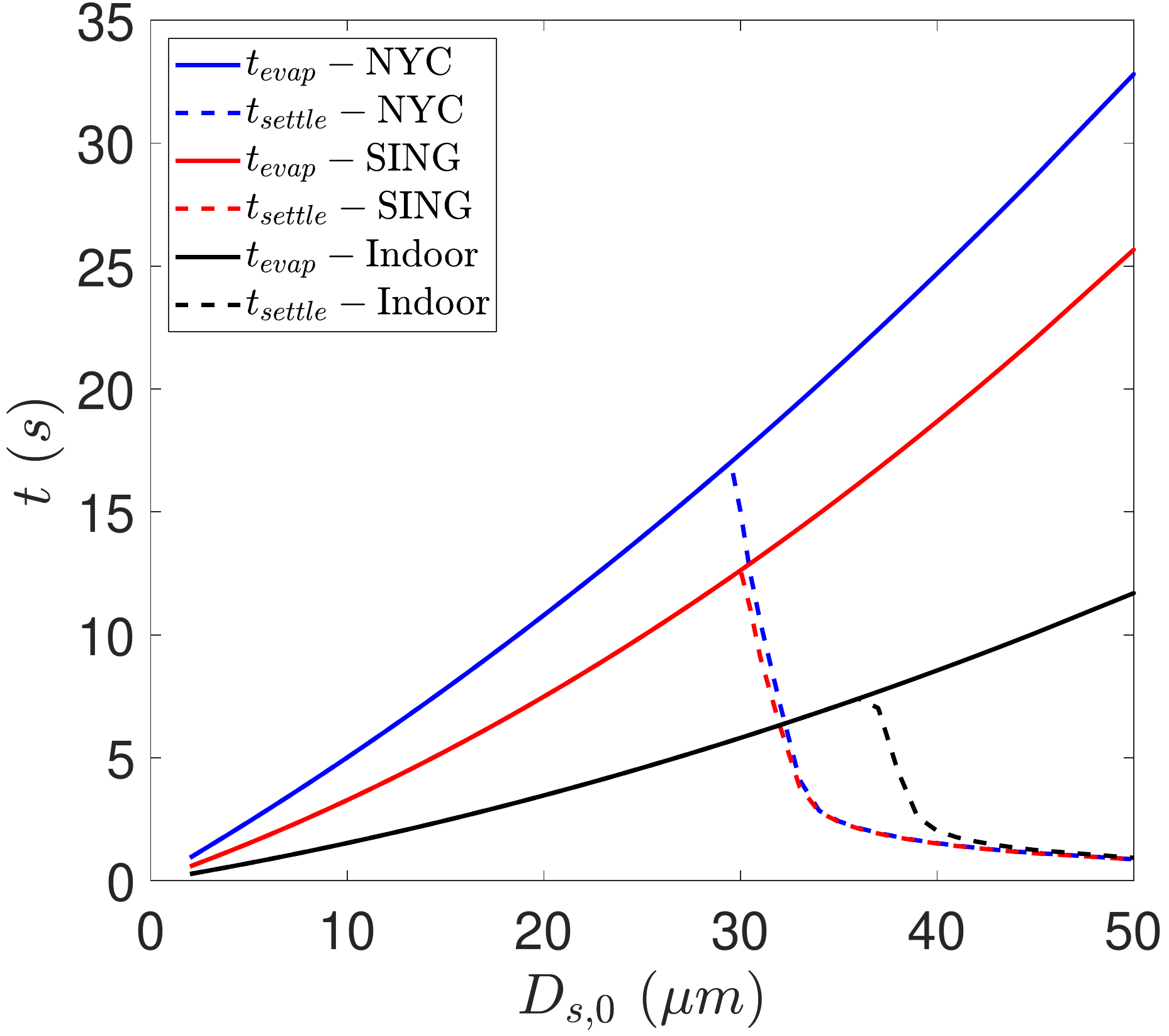}
\caption{Lifetime of droplets as a function of initial droplet diameter $D_{s,0}$. The solid line shows total evaporation time while the dashed line depicting settling time shows the time within which the droplets escapes the radius of collection $\sigma_H/2$.}
\label{fig:lifetime}
\end{figure}

\subsection*{Location specific droplet lifetime}
Next, we set out to use this model to predict the droplet evaporation characteristics at two very different geographical locations, specifically at New York City and at Singapore to finally connect it to the respective infection rates. The "outdoor" ambient conditions to be used are average ambient temperature and relative humidity at the two locations. We also consider "indoor" conditions, where temperature and relative humidity are controlled and are assumed to be same for both locations. 
To obtain the averaged outdoor ambient parameters, averaging time period for each location starts from five days prior to the date on which 10th Covid-19 positive case was reported at that location. The averaging ended five days before the time upto which model output was required. For New York City from February 25, 2020 to March 26, 2020, the temperature and relative humidity ranges are given by: $4.61^oC \leq T_{\infty} \leq 11.74^oC$ and $35.90\% \leq RH_{\infty} \leq 71.35\%$, respectively. The mean values that would be used for modeling New York City outdoor conditions are $T_{\infty} = 7.56^oC$ and $RH_{\infty}=55.28\%$. For Singapore, the time period spans from January 25, 2020 to March 18, 2020. The temperature and relative humidity in Singapore for that period ranges within  $25.67^oC \leq T_{\infty} \leq 31.85^oC$ and $58\%\leq RH_{\infty}\leq89.89\%$. The mean values that would be used for Singapore outdoor conditions are $T_{\infty}=28.17^oC$ and $RH_{\infty}=76.92 \%$. For indoor conditions we consider the standard values prescribed by American Society of Heating Refrigerating and Air-Conditioning Engineers for indoor air-conditioning \cite{ASHRAE}, i.e.  $T_{\infty}=21.1^oC$ and $RH_{\infty}=50 \%$.
Using these three sets of ambient conditions characterized by mean temperature and mean relative humidity at New York, Singapore and indoors, we estimate the droplet evaporation time over a range of droplet sizes $2 \mu m \leq D_{s,0} \leq 50 \mu m$. The final evaporation time of droplets of the above size range is shown in Fig. \ref{fig:lifetime}. Also, shown in the same figure is the settling time, which is calculated accounting for the decreasing diameter using the equation for Stokes settling velocity. 
\begin{equation} \label{Eqn:Stokes settling velocity}
w=(\rho_p - \rho_f)g D_s^2/18 \mu
\end{equation}
The settling time is estimated as that time by which the droplet gets out of the radius from which breathable air is collected - which has already been defined as $\sigma_H/2$ in Section 3. Mathematically $t_{settle}$ is obtained by Eqn. \ref{Eqn:Settling time} 
\begin{equation} \label{Eqn:Settling time}
\int_{0}^{t_{settle}} w dt = \sigma_H/2	
\end{equation}

In Fig. \ref{fig:lifetime}, $t_{settle}$ for different $D_{s,0}$ is plotted only when $t_{settle} \leq t_{evap}$. $t_{evap}$ being the overall evaporation time at the corresponding initial diameter of the droplet. From Fig. \ref{fig:lifetime} it is clear that for all conditions while $t_{evap}$ monotonically increases with $D_{s,0}$, $t_{settle}$ monotonically decreases with $D_{s,0}$ as expected. The purpose of this exercise is to estimate the maximum time an exhaled droplet can remain within the collection volume without being evaporated. This can be estimated by defining a characteristic droplet lifetime $\tau$, where

\begin{equation} \label{Eqn:Lifetime}
\tau = \max \Big\{ \min \big\{ t_{evap}, t_{settle} \big\}, \forall D_{s,0} \Big\}
\end{equation}

\begin{figure}
\centering
\includegraphics[width=0.8\textwidth]{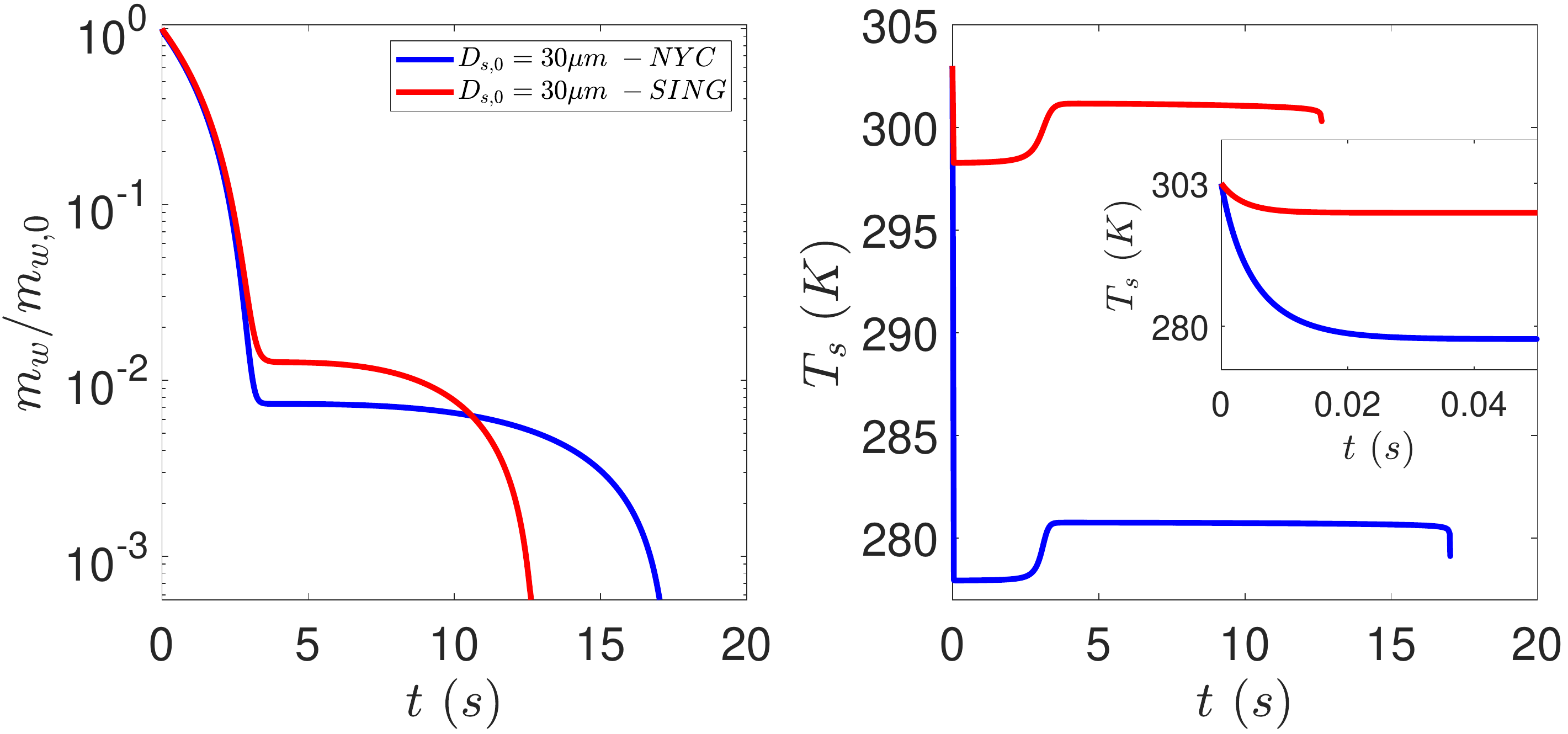}
\caption{Evolution of normalized mass of water in the droplet (left panel) and droplet temperature (right panel) as a function of temperature at the conditions of New York City and Singapore.}
\label{fig:normalzed mass and temperature at NYC and SING}
\end{figure}

From Fig. \ref{fig:lifetime} it can be found that $\tau$ at New York City and Singapore are $\tau_{NYC} = 17.02s$ while $\tau_{SING} = 12.63s$, respectively. The initial droplet diameter corresponding to outdoor conditions for New York City and Singapore are $D_{s,0,NYC}=29.5\mu m$ and $D_{s,0,SING}=30\mu m$, respectively. $D_{s,0,indoor}=36 \mu m$ for indoor conditions. Thus in Fig. \ref{fig:normalzed mass and temperature at NYC and SING} we look into the evolution of the normalized mass and temperature of the droplets at the two outdoor conditions. Fig. \ref{fig:normalzed mass and temperature at NYC and SING} clearly explains why  $\tau_{NYC}>\tau_{SING}$. We may recollect that the New York City outdoor ambient conditions considered corresponds to low temperature and low RH while Singapore outdoor condition corresponds to higher temperature and higher RH. Indeed higher RH at Singapore implies that the temporary hiatus in evaporation due to reduced vapor pressure is reached at a higher water mass load of the droplet, than at New York City condition. In both cases, this occurs at about 3 seconds. However, the higher temperature in Singapore condition results in faster crystallization kinetics due to the Arrhenius nature of the equation given by Eqn. \ref{Eqn:Crystal growth rate} which causes eventual faster crystallization rate, than at New York City condition. Figure \ref{fig:normalzed mass and temperature at NYC and SING}, clearly shows that although the knee in the solvent depletion profiles are attained at the same time, it is the crystallization and simultaneous desiccation dynamics that governs the eventual difference in lifetime of the droplets at two different locations. It remains to be seen whether this result holds for a detailed crystallization reaction mechanism. The temperature evolution plots shown in right panel of Fig. \ref{fig:normalzed mass and temperature at NYC and SING} also reveals how the droplet initially exhaled at $30^oC$ rapidly cools to the corresponding wet-bulb temperature to subsequently allow heat transfer into the droplet leading to evaporation. However, as the salt concentration reduces due to crystallization, the temperature rises subsequently, above the corresponding wet-bulb limits. 

\begin{figure}[ht!]
\begin{centering}
\includegraphics[trim=0 6cm 0 6cm 0,clip=true, totalheight=0.5\textheight,]{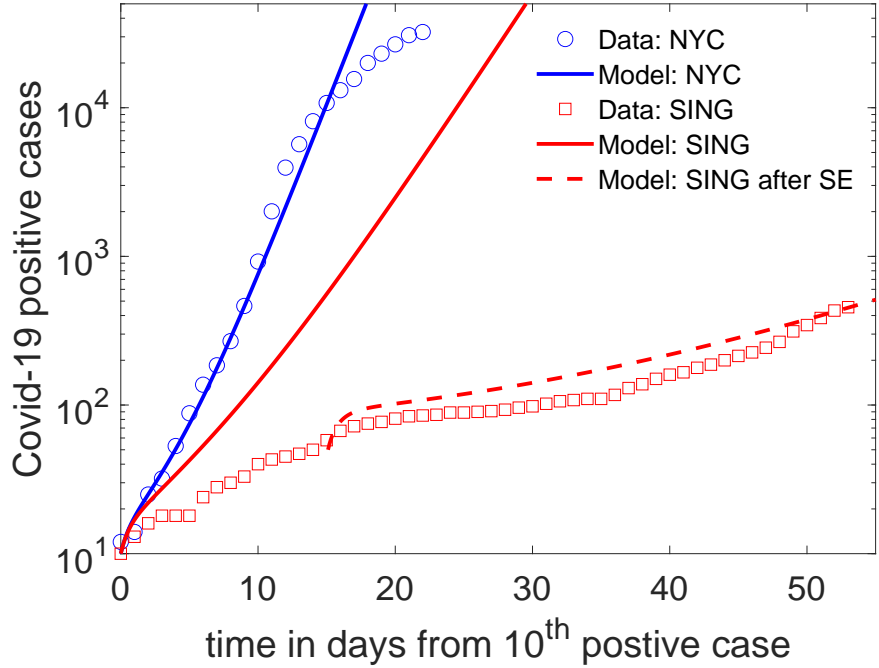}
\caption{Actual Covid-19 data and model output from Eqn. \ref{Eqn: Outdoor Indoor}. All parameters could be found in Table \ref{table:1}. Covid-19 data for New York City has been obtained from \cite{NYC} and data for Singapore is obtained from \cite{SIN}. SE implies Social Enforcement.}
\label{fig:NYC_SN}
\end{centering}
\end{figure}

\subsection*{Location specific growth rate of Covid-19 positive cases}
With the droplet lifetime available at the two locations, the corresponding infection rate constants given by Eqn. \ref{Eqn:k} could be evaluated. As such, the performance of the model could be tested using the time series data from New York City and Singapore and shown in Fig.~\ref{fig:NYC_SN}. All the parameters used to calculate $k_1$ to finally numerically calculate Eqn. \ref{P=f(t)} at the two different locations and four different conditions are described below in Table \ref{table:1}. Additionally, the average distance $X_{D,P}=\tau V_{D,P}$ travelled by
 the droplet cloud $D$ relative to $P$ is shown in the Table. $X_{D,P}$ ranges from 1.45m to 1.79m.
To account for infection at both outdoor and indoor, at each geographical location, Eqn. \ref{P=f(t)} can be modified as the following:

\begin{equation} \label{Eqn: Outdoor Indoor}
\begin{split}
[P]=[P]_0(1-\mathcal{F})(C_{1,o}e^{\lambda_{1,o}t} + C_{2,o}e^{\lambda_{2,o}t})  +[P]_0 (\mathcal{F})(C_{1,i}e^{\lambda_{1,i}t} + C_{2,i}e^{\lambda_{2,i}t}) \\
\implies N_P=N_{P,0}(1-\mathcal{F})(C_{1,o}e^{\lambda_{1,o}t} + C_{2,o}e^{\lambda_{2,o}t})  + N_{P,0} (\mathcal{F})(C_{1,i}e^{\lambda_{1,i}t} + C_{2,i}e^{\lambda_{2,i}t})
\end{split}
\end{equation}

\begin{table*}[h!]
\centering
\begin{tabular}{ |c|c|c|c|c| } 
 \hline
 Parameters   & NYC Outdoor & NYC Indoor & SING Outdoor & SING Indoor \\ 
 $n_{density} (people/km^2)$ & 10194 & 10194 & 8358 & 8358 \\ 
 $n_{total} (people/m^3)$ & 5.06x10$^{-4}$ & 5.17x10$^{-4}$ & 4.18x10$^{-4}$ & 4.23x10$^{-4}$ \\ 
 $D_{s,0} (\mu m)$ & 29.5 & 36.0 & 30.0 & 36.0 \\
 $T_{\infty} (K)$  & 280.76 & 294.44 & 301.17 & 294.44 \\ 
 $RH_{\infty} (\%)$ & 55.28 & 50 & 76.91 & 50 \\ 
 $\tau (s)$ & 17.02 & 7.41 & 12.63 & 7.41\\
  $t_c (s)$ & 5400 & 5400 & 5400 & 5400\\
 $V_{D,P} (m/s)$ & 0.11 & 0.20 & 0.13 & 0.20 \\ 
 $V_{DH}(m/s)$ & 2.905 & 2.996 & 2.931 & 2.996 \\ 
 $X_{D,P}$ & 1.79 & 1.45 & 1.66 & 1.45 \\ 

 $\sigma_{H}(m)$ & 0.124 & 0.124 & 0.124 & 0.124 \\ 
 $\sigma_{DH}(m)$ & 0.27 & 0.23 & 0.26 & 0.23 \\ 
$k_1 (day^{-1})$ & 0.985 & 0.339 & 0.548 & 0.278
 \\
$k_3 (day^{-1})$ & 1 & 1  & 1 & 1 \\
$k_3 (day^{-1})$ & 0.071 & 0.071 & 0.071 & 0.071 \\
$\mathcal{F}$ & 0.75 & 0.75 & 0.75 & 0.75 \\
$C_{1,o}$  & 1.178 &-& 1.351 &-         \\
$C_{2,o}$  &-0.178 &-&-0.351 &-       \\
$C_{1,i}$  &-&1.502 &-& 1.561        \\
$C_{2,i}$  &-&-0.502&-& -0.561        \\

$N_{P,0}$ & 10 & 10 & 10 & 10 \\
$\lambda_1 (day^{-1})$ & 0.544 & 0.195 & 0.325 & 0.154 \\
$\lambda_2 (day^{-1})$ & -1.616 & -1.266 & -1.396 & -1.226 \\
\hline
\end{tabular}
\caption{Parameters to calculate Eqn. \ref{Eqn: Outdoor Indoor} at ambient conditions corresponding to New York City (NYC) and in Singapore (SING) averaged over the specified time periods. $n_{density}$ (units $people/km^2$) is converted to $n_{total}$ (units $people/m^3$) using $n_{total} = n_{density}/[10^6 \times (1.8 + \sigma_D/2)]$. $k_1$ is calculated using Eqn. \ref{Eq:k_1 equation}. $t_c$, the average interval between expiratory events (coughing and sneezing) is calculated assuming an average person coughs 12 times and sneezes 4 times within 24 hours \cite{munyard1996much, hansen2002often}.}
\label{table:1}
\end{table*}

Here, $\mathcal{F}$ denotes the fraction of the infectious population that infected others, at indoors, at time $t=0$. We assume $\mathcal{F}=0.75$ that is initially, most infections occurred indoors. $\lambda_{1,o}, \lambda_{2,o}$ and $\lambda_{1,i}, \lambda_{2,i}$ denote the eigenvalues from the system of ODEs given by Eqn. \ref{Eqn:dPdt0} for outdoor and indoor $k_1$, respectively. $N_P$ and $N_{P,0}$ are the total number of infectious people at day $t$ and at day $t=0$, respectively, in a location for a given population density $n_{total}$ and ambient conditions. The constants $C_1, C_2$ are determined from the eigenvectors of the $k$-matrix shown in Eqn. \ref{Eqn:dPdt0} and the initial conditions. We assume $N_{P,0}=N_{P^*,0}$.
Figure \ref{fig:NYC_SN} shows that the model qualitatively conforms to the available data from New York City (NYC) and Singapore (SING). Though the match between NYC data and model output is excellent, given the number of parameters involved, we believe that such exact match could be fortuitous. However, we indeed find that the large difference in the Covid-19 positive growth rates between two highly populated cities could be partly attributed to their difference in outdoor ambient temperature and relative humidities. However, the actual growth rate in Singapore while initially following the model, dropped much further with time. This could be attributed to the large scale enforcement of temperature check, contact tracing, quarantining and widespread usage of masks at very early stages of the outbreak. If we assume that social enforcement measures were implemented early in Singapore \cite{ng2020evaluation} such that concentration of susceptible healthy individuals reduced to $[H]_{SE}=0.3$ after the 50th case was reported, then utilizing Eqn. \ref{Eq:Social Enforcement} and same $k_1$ as before, we find that the model follows the actual data rather closely. Thus, according to the model the comparatively slower growth rate of the outbreak in Singapore is attributed to both favorable weather conditions as well as appropriate social enforcement measures.

\section*{Summary} 
Respiratory flow ejected by human beings consists of a polydisperse collection of droplets. Since the virus is an obligatory parasite, it needs an intra-cellular environment to proliferate. However, if rapid desiccation of fluid from the droplets shrink the volume of the bio-material and drastically prevent the virus from multiplying, a relatively sparse viral load to any non-affected person is presented. 

\begin{figure}
\begin{centering}
\includegraphics[width=0.8\columnwidth]{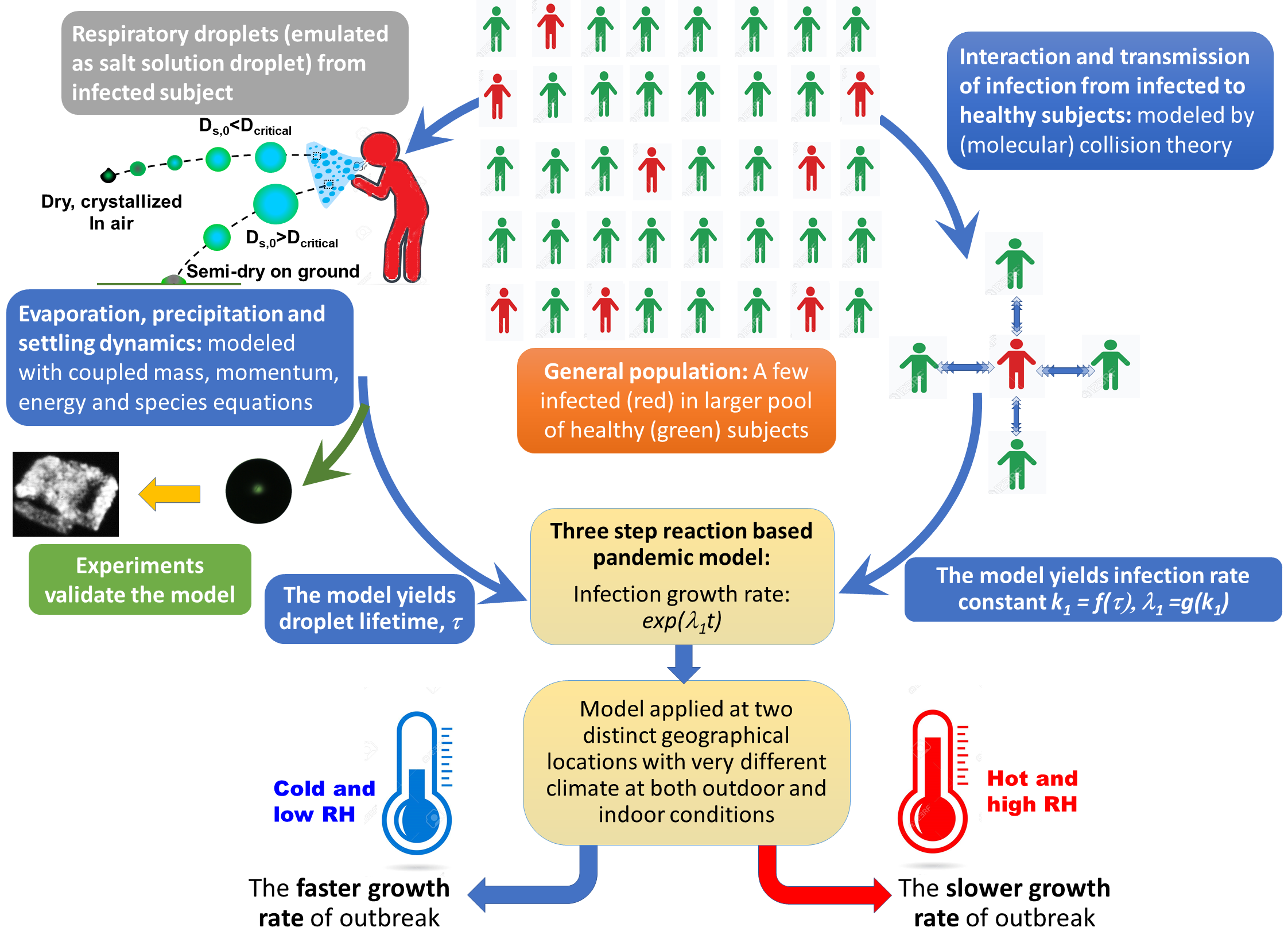}
\caption{Flow-diagram outlining the interconnections of the model developed}
\label{fig:flowchrt}
\end{centering}
\end{figure}

In this paper we have presented a model for the early phases of Covid-19 outbreak as a function of ambient temperature and relative humidity, at both outdoor and indoor conditions. The model and its inter-dependencies on the different physical principles/sub-models is summarized in Fig. \ref{fig:flowchrt}. It must be recognized that the model assumes conditions where transmission occurs solely due to inhalation of infected respiratory droplets alongside many other simplifying assumptions. After being ejected, smaller droplets attain the wet bulb temperature corresponding to the local ambience, and begins to evaporate. However, due to presence of dissolved salt the evaporation stops when the size of the droplet reaches about 20-30\% of the initial diameter. But now, the droplet salt concentration has increased to levels that trigger onset of crystallization. The evaporation time of the salt solution droplet is a however a complex manifestation of the intricate interplay between several physico-chemical phenomena: droplet cooling, evaporation, increasing salt concentration leading to crystallization. Of course, the entire process competes with settling - the process by which larger droplets fall away before they can evaporate. The smaller of the two: evaporation time and settling time thus dictates droplet lifetime $\tau$. In any case, the infection rate constant derived using collision theory of reaction rates, is shown to be a function of the respiratory droplet lifetime ($\tau$). Since the infected population size depends almost exponentially on this rate constant, the model finds that given similar relative humidity between two places, infection spreads more rapidly in lower temperature conditions than in their warmer counterparts. While qualitative trends of actual Covid-19 data are predicted by the model, it is found that difference between outdoor ambient conditions - temperature and RH cannot quantitatively explain the difference between data obtained for two different metropolis. It is found that inclusion of social enforcement measures could account for the remaining difference. Thus, it can be concluded (within limitations of the model and several associated assumptions) that while warm weather indeed can lead to slower growth rates, strict social enforcement - including but not limited to social distancing, contact tracing, quarantining and population wide usage of non-medical masks that can prevent entry of droplets of size 10-30$\mu m$ into the respiratory tract, is still much required to control the Covid-19 pandemic.

\section*{Acknowledgement}
The authors thank Prof. C. K. Law, Prof. D. Lohse and Prof. K.N. Lakshmisha for reading the paper and for providing valuable comments. The authors thank Dr. Chaitanya Rao, Dr. Lalit Bansal and Mr. Sandeep Hatte for independently checking some of the calculations.


\bibliography{pnas-sample}{}
\bibliographystyle{ieeetr}

\end{document}